\newcommand{\Teff}{$T_{\rm eff}\,$}        
\newcommand{\DYDZ}{$\Delta$$Y$/$\Delta$$Z\,$}        
\newcommand{\Zsun}{$Z_\odot\,$}        
\newcommand{\Msun}{$M_\odot\,$}        
\newcommand{\Menv}{$M_{env}\,$}        
\newcommand{\Mtot}{$M_{tot}\,$}        
\newcommand{\Mcore}{$M_{core}\,$}        
\documentstyle[11pt,aaspp4,psfig]{article}
\received{22 October 1996}
\slugcomment{ApJ, vol. 482, in press}

\lefthead{Yi et al.}
\righthead{UV Bright Phase of Metal-Rich Horizontal-Branch Stars}
 
\begin{document}

\title{On the UV Bright Phase of Metal-Rich Horizontal-Branch Stars}
\author{Sukyoung Yi\altaffilmark{1}}
\affil{Department of Astronomy, Yale University, P.O. Box 208101, New Haven, CT06520-8101, NASA/Goddard Space Flight Center, Code 681, Greenbelt, MD 20771\\ email: yi@shemesh.gsfc.nasa.gov}
 
\author{Pierre Demarque and Yong-Cheol Kim}
\affil{Department of Astronomy, Yale University, P.O. Box 208101, New Haven, CT06520-8101 \\ demarque@astro.yale.edu, kim@astro.yale.edu}

\altaffiltext{1}{National Research Council Research Associate.}

\begin{abstract}

   We consider the origin of the UV bright phase of metal-rich helium-burning 
stars, the slow blue phase (SBP), that was predicted by 
various earlier works.
   Based on improved physics including OPAL opacities, which is the same
physics as was used in the construction of the new Yale Isochrones, 
we confirm the
existence of the SBP.  In addition to our grid of evolutionary tracks,
we provide an analytical understanding of the main characteristics of the
SBP phenomenon.

   The SBP is slow because it is a slow evolving helium-shell-burning phase
which is analogous to the early asymptotic giant branch phase.
   The SBP of a more metal-rich star is slower than a metal-poor counterpart
if their \Teff's are the same because a more metal-rich
helium-burning star has a smaller mass than a metal-poor one and
because lifetime increases as mass decreases.

   Metal-rich helium-burning stars easily become hot because the 
luminosity from the hydrogen-burning shell is extremely sensitive 
to the mean molecular weight $\mu$ whereas the luminosity from the 
helium-burning core is not.
   Under the assumption of a positive \DYDZ, helium abundance plays the most 
important role in governing $\mu$, and thus Dorman and collaborators found 
that the SBP occurs only when $Y \gtrsim$ 0.4 when \DYDZ $\gtrsim$ 0.
   We suggest that the SBP phenomenon is a major cause of the UV upturn 
phenomenon in giant elliptical galaxies as will be shown in subsequent
papers.
   The new HB tracks can be retrieved from S.Y.'s web site 
http://shemesh.gsfc.nasa.gov/astronomy.html.
\end{abstract}


\section{Introduction}

   Early numerical studies of the advanced evolution of metal-rich stars
by Demarque and Pinsonneault (1988) and by Horch, Demarque, \& Pinsonneault 
(1992) suggested that under the simplest assumptions about mass loss on the 
red giant branch (RGB) and galactic helium enrichment (\DYDZ $\approx$ 2 -- 3),
stars in the core helium-burning (i.e. post-RGB) phase of evolution do not 
evolve into asymptotic giant branch (AGB) stars. 
   Instead, before becoming white dwarfs (WDs), they evolve into UV bright 
objects.  
   Horch et al. (1992) named this evolutionary phase, the slow blue phase 
(hereafter SBP).
   Horch et al. (1992) also found that the transition total mass,
$M_{tot}^{tr}$, the total mass below which the UV bright phase occurs, is 
highly sensitive to metallicity in the sense that it increases with increasing 
metallicity.
   In other words, a larger fraction of stars become UV bright instead of
becoming post-AGB (PAGB) stars as metallicity increases.

   These UV bright SBP stars can be divided into two groups.
   Those in the first group are very low-mass helium-burning stars that never 
become red (i.e. AGB) but instead evolve directly into the WD phase
(Sweigart, Mengel, \& Demarque 1974).
   They spend all of their lifetimes on the hot horizontal branch 
and more luminous hot phases.
   Those in the second group are moderately low-mass helium-burning stars
that first evolve to become AGB stars but then quickly return to the SBP
without reaching the tip of the AGB.
   Both types are sensitive to metallicity in the sense that
a larger fraction of helium-burning stars become SBP stars as metallicity
increases.
 
   Greggio \& Renzini (1990) postulated the existence of the first group
in metal-rich systems
using a gedanken experiment and named it the AGB-manqu\'{e} stage.
   Meanwhile, Castellani \& Tornamb\'{e} (1991) independently noticed the 
second group in their calculations and called it the post-early AGB (P-EAGB).
   The SBP phenomenon was later confirmed by more detailed numerical 
calculations (\cite{hdp92}; \cite{dro93}; \cite{fag94}).

   These discoveries are provocative because they are opposite
to the traditional concept of the evolution of HB stars as a function
of metallicity (more metal-rich HB stars are cooler).
   They suggest that old, metal-rich stellar systems may be in fact better 
UV light generators than metal-poor systems if they are sufficiently metal-rich
($Z \gtrsim$ \Zsun) to experience the SBP phenomenon.
   Moreover, these discoveries are consistent with the hitherto mysterious
UV flux-metallicity correlation (UV upturn phenomenon) in giant elliptical 
galaxies (\cite{cw79}; \cite{b88}).
   That is, if such metal-rich, UV bright HB stars are indeed the major
UV sources in giant elliptical galaxies, the observed positive 
UV upturn-metallicity relationship can be naturally understood.
   
   In the same direction, another important step forward was made by 
Castellani and Castellani (1993), who pointed out that the inclusion of mass 
loss can have an effect on the evolution near the tip of the giant branch, 
particularly for stars with low envelope masses.  Thus if the mass
loss takes place primarily near the giant branch tip, the core contraction 
that triggers the helium flash can be initiated even if the
helium core mass is below the critical mass for a no-mass loss model.
This results in the possibility of helium ignition as the star reaches
the extended horizontal-branch, and for subdwarf B core 
masses smaller than the core masses of ordinary horizontal branch 
stars.  This work was recently followed up and extended by D'Cruz et al.
(1996), who explained the existence of sudwarf B stars by postulating
extreme mass loss on the giant branch.

   However, physics of such late stellar evolution is not well-understood
and is somewhat sensitive to the input physics, such as opacities. 
   Moreover, because of difficulty in finding the actual stars, owing 
to their rarity in stellar samples, some doubt  has been cast
about this purely theoretical\footnote{There may have been several
empirical discoveries of such stars recently. The most promising examples
are the hot stars in NGC\,6791, an old, metal-rich open cluster
(\cite{ku92}; Liebert, Saffer, \& Green 1994; \cite{kr95}).} 
prediction (\cite{l94}).

   Motivated by these theoretical discoveries, we have carried
out extensive calculations of advanced stellar evolution for a variety of
chemical compositions and masses in order to investigate the validity
of the theoretical predictions and understand the physical basis of
the SBP phenomenon better.
   
\section{Construction of Evolutionary Tracks}

   The same improved physics that has been used in the calculations of the
new Yale Isochrones 96 (\cite{d96}) has been used 
to construct the helium-burning phase evolutionary tracks.
It includes OPAL opacities (\cite{ri92}), Kurucz low temperature 
opacities (\cite{k91}), and improved energy generation rates (\cite{bp92}). 

   Semiconvection is included, but overshooting, diffusion, and the
Debye-H\"{u}ckel correction have not been taken into account in order to be 
consistent with the MS through RGB calculations in the isochrones. 
   The neutrino cooling rates used on the giant branch, which are 
important in determining the helium core mass at helium ignition at 
the tip of the RGB, are those of Itoh et al. (1989).
   These physical considerations have an influence particularly on the 
evolutionary time scale of the model, which are still uncertain
by several gigayears (Chaboyer and Kim 1995; Chaboyer et al. 1996a).  

   For instance, the age scale adopted by Chaboyer et al. (1996a) in 
their Monte Carlo study is 17\% lower than the one used in the study of the 
galactic halo chronology recently published by Chaboyer et al. (1996b).  
   This difference can be accounted for in the following way: a 7\% decrease
due to the improvements in the equation of state mentioned above, and 
another 7\% decrease due to the inclusion of helium diffusion.  
   The additional 3\% decrease is due to an upward revision in 
the $[\alpha/Fe]$ ratio from 0.4 to 0.55 in Chaboyer et al. (1996a). 
   Note that the sensitivity of globular cluster ages to $[\alpha/Fe]$
is modest, a result in agreement with the earlier conclusion of 
Chieffi et al. (1991).   
   Therefore, considering other unknown sources of uncertainty, only relative 
age is meaningful in this study.
   However, the general evolutionary patterns are not sensitive to
these details in the input
physics. Therefore all the conclusions that will be made in this paper
remain the same except the absolute sense of the age.

   Metal-rich ($Z \gtrsim 0.01$) models have been constructed for 
\DYDZ = 2 and 3. 
   It is believed that this range embraces the true value, considering that 
\DYDZ = 2.75 for the Sun, assuming primordial chemical composition 
($Z$, $Y$)$_0$=(0, 0.23) and current composition ($Z$, $Y$)$_{\odot}$ = 
(0.0188, 0.2817), respectively (\cite{ag89}; \cite{gue92})\footnote{ 
There is still some uncertainty in the solar helium abundance which, when one 
allows for Coulomb interactions in the equation of state and for 
the effects of helium diffusion, would be closer to $Y = 0.27$ (Proffitt 1992; 
Guenther, Kim, \& Demarque 1996).}. 
    The galactic helium enrichment has not been taken into account for 
metal-poor ($Z < 0.01$) stellar evolution because the effect is 
negligible. Figures 1 -- 2 show the evolutionary tracks, and the
detailed information of the models is given in Table 1.

   Fewer models than actually constructed are displayed in Figures 1 -- 2
for the sake of clarity. As generally accepted, more massive stars are cooler
until the total mass $M_{tot}$ reaches a certain value (about 1 \Msun),
but above that mass the trend becomes reversed. 
   For instance, the most massive model, 
$M_{tot} = 1.5 M_{\odot}$ is hotter than the next massive model,
$M_{tot} = 0.9 M_{\odot}$, which is clearly seen in the $Z = 0.0001$ panel
(\cite{dh75}).
   It is also seen that the SBP (slow blue phase), the UV bright phase, 
becomes much more prominent as metallicity increases.

   To demonstrate the effect of the 
improved physics, some new models are compared in Figure 3 with models 
that were constructed using the old physics. 
   As expected, the new metal-rich models with improved opacities  
are redder than the old models. 
   On the other hand, the new metal-poor models, which are less sensitive 
to opacities, are slightly brighter than the old ones due to
recent revisions to the nuclear energy generation rates.

    While the evolutionary tracks are generally similar,
some tracks show a notable difference.
For example, the new 0.56 \Msun model in the left panel of Figure 3
becomes an AGB star whereas the old model becomes a SBP star. 
    Such a difference
is present only for metal-rich models where the change in the opacities
is significant. In addition, such an effect is more outstanding
near the transition total mass, $M_{tot}^{tr}$ (a total mass below 
which the UV bright phase occurs), where a subtle difference in 
physics easily alters the fate of the star. 
    By and large, the net effect is that we expect fewer
UV bright stars if we use the new calculations based on improved physics.

\section{UV Bright Phase in the Metal-Rich Systems}

\subsection{Why is the SBP Slow?}

   The question of the origin of the SBP can be divided 
into two parts: (1) why is the SBP slow? and (2) why is it blue?
   The first question is easier to answer.
   First, the SBP is a  helium-shell-burning phase like early AGB phase in
which stars evolve very slowly.
  Moreover, the lifetime of a star in this slow phase (from the ZAHB to evolved
HB phase) increases with metallicity.
   This can be understood easily based on the so called mass-luminosity
relation: a more massive star is brighter and thus dies more quickly.
   Similarly, a core helium-burning star with a smaller \Mcore lives longer
than the one with a larger \Mcore when the total masses are the same.

   To begin with, a metal-rich red giant experiences the helium core flash
before its core becomes as large as that of a metal-poor counterpart
(\cite{sg78}).
   This is because, in a more metal-rich red giant, the 
higher opacity outside the degenerate core causes the temperature in the core 
to rise to that required to initiate the helium core flash more quickly.
   So, a more metal-rich core helium-burning star has a smaller \Mcore
than a less metal-rich one.
   Therefore, a more metal-rich model has a longer lifetime in the core
helium-burning phase.  
   For example, let us compare two core helium-burning stars with the
same \Menv but with different metallicities.
   A model of (\Mtot = 0.445 \Msun, $Z$ = 0.06, $Y$ = 0.41,
\Menv = 0.005 \Msun, log \Teff (ZAHB) = 4.38) has a lifetime of 270 Myrs in
the core helium-burning phase before it becomes a WD, which is 80\%
longer than the lifetime of a metal-poor model (150 Myr) of
(0.495, 0.004, 0.24, 0.005, 4.42).
   As a result, when we have two stellar systems that have similarly hot HBs
but different metallicities as the examples given above, then the
metal-rich system would have twice as many hot HB stars as the metal-poor
system does.

\subsection{Why is the SBP Blue?}   

   The second question is more difficult to understand.
   The classical understanding about the SBP (\cite{hdp92}) is as follows. 
A core helium-burning star with a thin envelope burns up its 
hydrogen burning shell quickly compared to the core helium-burning time scale,
giving the star little time to expand the envelope slowly and become an
AGB star. 
   Instead, it becomes a hot, UV bright star because it shrinks in radius
(i.e. surface gravity increases) as its hydrogen burning in the shell
quickly decreases with time.  
   In order to see this phenomenon, the envelope mass, $M_{env}$, should 
be as low as $M_{env} \lesssim$ 0.05 \Msun for a star with $Z$ = \Zsun.
   It requires many billions of years for a stellar population to develop
a substantial fraction of HB stars with such a low-mass envelope,
assuming a moderate mass loss efficiency.
   Horch et al. (1992) suggested that this UV bright phase should
occur even for  stars with a larger total mass if their metallicity is higher. 
   They suggested that the transition total mass, $M_{tot}^{tr}$, 
increases monotonically with increasing metallicity. 
Therefore even a star with a massive envelope may develop a UV bright star.
   Dorman et al. (1993) also found the same phenomenon and emphasized 
that this phenomenon becomes efficient only when $Y \gtrsim$ 0.4. 

   A more detailed and quantitative analysis has been carried out in this
study in order to check the suggestions that were made by 
the previous theoretical predictions, and to understand the cause of the 
UV upturn phenomenon better.
    First, Figures 1 -- 2 qualitatively but clearly show (1) the  
SBP occurs only in the very low-mass core helium-burning stars, 
and 
(2) it happens to the more massive stars when the metallicity is higher
under the assumption of a positive \DYDZ.

    The UV bright phenomenon is illustrated in Figure 4. 
The left panel is the relation between the transition total mass, 
$M_{tot}^{tr}$, and metallicity. It shows that $M_{tot}^{tr}$ is not 
sensitive to metallicity at all for $Z <$ \Zsun and becomes very 
sensitive especially when $Y \gtrsim$ 0.4 as Dorman et al. (1993) suggested.
However, a more realistic comparison is the relationship between the 
transition envelope mass ($M_{env}^{tr}$) and metallicity because the 
classical understanding of the SBP
requires a small $envelope$ mass, not a small $total$ mass. 
    The dependence of $M_{env}^{tr}$ on metallicity is shown 
in Figure 4-(b) in which $M_{env}^{tr}$ monotonically
increases as a function of metallicity for the positive values of \DYDZ.
    
   It is important to understand why such a tight correlation between
$M_{env}^{tr}$ and metallicity exists.
To begin with, it is clear that chemical composition plays an important role   
in governing $M_{env}^{tr}$. We would like to know what aspect of
metallicity is the main element of this behavior. 
   Therefore the evolutionary tracks of the metal-rich models with the same 
mass, but with different chemical composition and/or core mass have been 
compared with one another, in order to disentangle the effects of one 
variable from those of another.

    Figure 5 shows the evolutionary tracks of eight models with different
set of parameters. The range of variation in the input parameters is 
consistent  with current expectations. 
    It is apparent that only the models with higher $Y$ become UV 
bright. An increase in $Z$ from 0.06 to 0.1 does not play as important
a role in making a star evolve to be a UV bright star as an increase
in $Y$ from 0.35 to 0.43 does. 
   This agrees with Dorman et al.'s (1993) suggestion that the SBP occurs
only when $Y \gtrsim 0.4$.
   Certainly the core mass plays little role within this range.

   Figure 6 confirms Horch et al.'s (1992) suggestion;
the UV bright phase is caused by the fast exhaustion of the hydrogen burning
shell due to the high sensitivity of $L_{H}$ (luminosity 
from the hydrogen burning) to $\mu$ (mean molecular weight). 
    Those models that become UV bright stars burn up hydrogen in the CNO 
cycle much more efficiently than the others, as shown in the top panel
of Figure 6.

   A simple analytic relationship between luminosity $L$ and $\mu$ in the
standard model was well described in many early works including Clayton
(1968, Eqn. 3-190). This formalism was originally developed for
the stars with single energy source. Therefore, the exact validity for
core helium-burning stars (with double energy sources) is debatable.
We assume that such a simplified model should be reasonable if it is used 
only to demonstrate the relation between $L$ and $\mu$ approximately.
According to such a simplified model, luminosity in Clayton's Eqn. 3-190
can also be expressed as follows;
\begin{equation}
L \propto M^{3} \, (\,\mu\,\beta_{c})^{4} \, <\,\kappa\,\eta_{n}\,>^{-1} 
\end{equation}
where $\beta_{c}$ is the ratio of gas pressure to total pressure at the 
stellar center, $\kappa$ is opacity, $\eta_{n}$ is the ratio of the average 
rate of nuclear energy generation interior to a certain point to the average 
for the whole star, and $M$ is the total mass of the star.
 
   Eqn. (1) explicitly shows that the luminosity depends on the 
source of the opacities. 
Since the major opacity sources corresponding to the temperature and density
of the helium-burning core and the hydrogen-burning shell are electron 
scattering ($\kappa \propto 1\,+\,X$) and free-free absorption 
($\kappa \propto \rho \, T^{-3.5}$), respectively, an analytic form of the
luminosity from the hydrogen burning, $L_{H}$, and the luminosity from the
helium burning, $L_{He}$, can be easily derived using a different absorption 
formula.
\begin{equation}
L_{H} \propto \mu^{7.5}
\end{equation}
\begin{equation}
L_{He}\propto \mu^{4}
\end{equation}
 
   In the hydrogen-burning shell, the mean molecular weight, $\mu$,
\begin{equation}
\mu = ( 2X + 0.75Y + 0.5Z )^{-1},
\end{equation}
can be approximated using the surface chemical composition based on the 
assumption of a full ionization because the hydrogen-burning region is very 
hot.
   On the other hand, the same expressions cannot be used for the $\mu$ in the
helium core because the core is mostly (in general, $Y_{\rm c} \gtrsim 0.9$)
composed of helium and a little bit of heavier elements (mostly carbon) 
regardless of the surface chemical composition.
   Therefore, the $\mu$ in the Eqn. (3) becomes more or less constant 
causing $L_{He}$ to be insensitive to the change in chemical compositions.
   Consequently, contrary to a simple conjecture, increase in helium abundance
increases $L_{H}$ significantly while it has little effect on $L_{He}$
as shown in Figure 6.
   Figure 6 shows that, in the beginning of the core helium-burning phase,
$L_{He}$ is not affected by the change in chemical compositions much,
whereas $L_{H}$ is sensitive to helium abundance.
   Once evolution begins, even $L_{He}$ increases as $Y$ increases because
of the inward effect from the higher $L_{H}$ in the hydrogen-burning shell.

   Indeed, Figure 7 shows that $M_{env}^{tr}$ is a sensitive function of 
$\mu$. However, it should be noted that chemical composition has other 
complicated second order effects on the evolution besides the effects on 
luminosity. Thus, the simple analytic formulae, that are originally
developed to work for MS stars, are not always precisely true 
in the practical numerical calculation. This simplified analytical
approach should be only demonstrating the relationship between luminosities
and the mean molecular weight, $\mu$, which helps us understand the origin 
of the UV bright phase of metal-rich stars. 

\section{Summary}
  
    The earlier prediction of the UV bright phase of helium-burning stars,
the slow blue phase (SBP), has been confirmed in this study based on 
improved physics. 
   According to this calculation, the SBP is more easily achieved when
helium abundance is higher because stars richer in helium burn up their
hydrogen-rich envelopes faster. Under the assumption of a positive
galactic helium enrichment (\DYDZ), this means that more metal-rich stars
become UV bright SBP stars more easily.
   This confirms the results of both Horch et al. (1992) and Dorman et al. 
(1993). 

    The SBP, an intrinsically slow evolving phase analogous to early
AGB phase, becomes slower as metallicity increases because more metal-rich
helium-burning stars are less massive than the less metal-rich counterparts
and because lifetime increases as mass decreases.
   Metal-rich helium-burning stars easily become hot because of
the different sensitivity of the luminosity from the hydrogen-burning shell 
and the helium-burning core to the mean molecular weight $\mu$. 
   Under the assumption of a positive \DYDZ, helium abundance, which
plays the most important role in governing $\mu$, has a dominant effect, 
and thus Dorman et al. (1993) found that the
SBP occurs only when $Y \gtrsim$ 0.4 when \DYDZ $\gtrsim$ 0.
   In principle, an extremely metal-rich hypothetical star may become
a SBP star even if the helium abundance is not that high since the SBP  
is not a direct function of $Y$ but a function of $\mu$.
   However, whether such hypothetical stars exist is questionable.
   On the contrary, if \DYDZ is higher than we assumed in this study, namely
\DYDZ $>$ 3, stars do not have to be very metal-rich in order to experience
the SBP phenomenon.
   Existing empirical data make us believe that the true \DYDZ should be
either within the bracket of 2 -- 3, or at least close to it.
   A more accurate determination of \DYDZ is required.
   
   The HB tracks were constructed using the same input physics that is used
to construct the new Yale Isochrones.
   These new HB tracks are qualitatively consistent with the previous
models (e.g. \cite{hdp92}; \cite{dro93}), but slightly different mainly
because of the improvement in opacities.
   When the mass of the star is near the transition mass, new models
tend not to evolve into the SBP.
   Therefore, those who are using Yale Isochrones in their galaxy or star 
cluster modelings are strongly recommended to use the HB tracks listed in this 
study\footnote{Both the Yale Isochrones 1996 used in this study and 
the HB tracks can be retrieved from S.Y.'s web site 
http://shemesh.gsfc.nasa.gov/astronomy.html.}.
   
   Some elements in the stellar evolution theory are still uncertain although
general qualitative evolutionary phenomena would not be affected by such
details. Thus only the relative ages should be taken seriously. 
   In case one adopts the final result (e.g. galaxy models from population 
synthesis based on stellar evolution theory), the ages of the models should 
be renormalized with respect to the ages of the oldest stellar systems 
(e.g. Galactic globular clusters).

   Whether the SBP phenomenon is mainly responsible for the UV upturn 
phenomenon in giant elliptical galaxies will be investigated in the
following papers (\cite{ydo96a}; \cite{ydo96b}).

\acknowledgements

   We thank Allen Sweigart, Augustus Oemler, Richard Larson, and Harry 
Ferguson for constructive comments. We are also grateful to the anonymous 
referee for many useful comments and clarifications.
   This work was a part of the Ph.D. study of S.Y. (\cite{yi96}) and was 
supported in part by NASA grants
NAGW-2937 (S.Y.), NAG5-1486 and NAG5-2469 (P.D. and Y.-C.K.).

\clearpage

\begin{table*}
\caption{Information for the models in Figures 1 \& 2.} \label{tbl-1}
\begin{center}
\begin{tabular}{rrrc}
\tableline
\tableline
$Z$   &  $Y$   &     $M_{core}$(\Msun)             &$M_{env}$(\Msun)\tablenotemark{\dagger} \\
\tableline
0.0001 &0.236 &0.501  &0.004  0.02  0.06  0.10  0.14  0.22  0.40  1.00 \\
0.0004 &0.237 &0.498  &0.007  0.02  0.06  0.10  0.14  0.22  0.40  1.00 \\
0.0010 &0.242 &0.488  &0.005  0.02  0.04  0.07  0.11  0.15  0.41  1.01 \\
0.0040 &0.241 &0.490  &0.007  0.02  0.04  0.07  0.11  0.15  0.41  1.01 \\
0.0100 &0.250 &0.470  &0.005  0.02  0.03  0.05  0.09  0.43  1.03 \\
0.0100 &0.260 &0.470  &0.005  0.02  0.03  0.05  0.09  0.43  0.103 \\
0.0200 &0.270 &0.470  &0.005  0.02  0.03  0.05  0.09  0.43  1.03  \\
0.0200 &0.290 &0.460  &0.005  0.02  0.04  0.06  0.10  0.44  1.04  \\
0.0400 &0.310 &0.460  &0.005  0.02  0.04  0.06  0.10  0.44  1.04 \\
0.0400 &0.350 &0.450  &0.005  0.02  0.04  0.07  0.11  0.45  1.05 \\
0.0600 &0.350 &0.450  &0.005  0.02  0.05  0.07  0.11  0.45  1.05 \\
0.0600 &0.410 &0.440  &0.005  0.02  0.06  0.08  0.16  0.46  1.06 \\
0.1000 &0.430 &0.440  &0.005  0.02  0.05  0.12  0.16  0.20  0.46  1.06 \\
0.1000 &0.530 &0.430  &0.005  0.02  0.05  0.09  0.17  0.29  0.47  1.07 \\
\tableline
\tableline
\end{tabular}
\end{center}
\tablenotetext{\dagger}{envelope mass: $M_{env} \equiv M_{tot} - M_{core}$}
\end{table*}

{}
\clearpage

\begin{figure}
\plotone{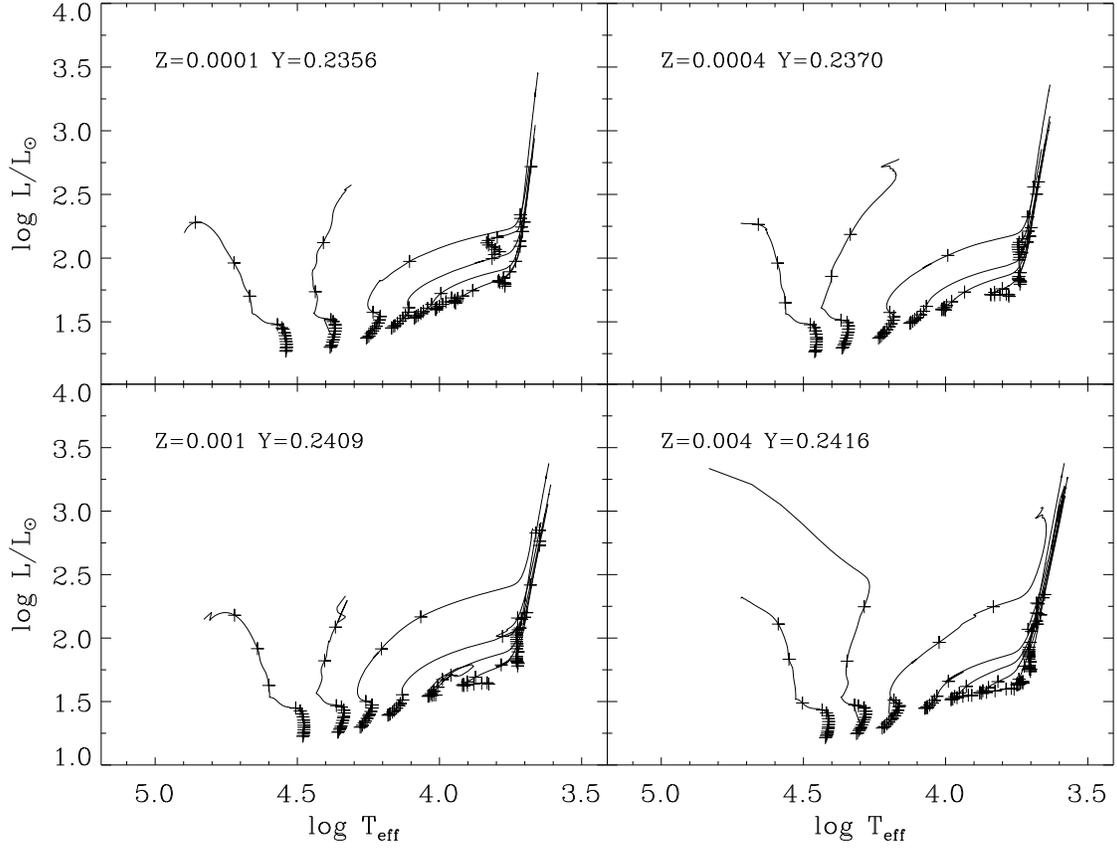}
\caption{Evolutionary tracks of metal-poor, core 
helium-burning stars in the theoretical CMD.
Each plus sign denotes 10 million years. Only very low-mass stars become UV
bright, and their envelope mass is very small. The details of the
models are listed in Table 1. \label{fig1}}
\end{figure}

\clearpage
\begin{figure}
\epsscale{0.8}
\plotone{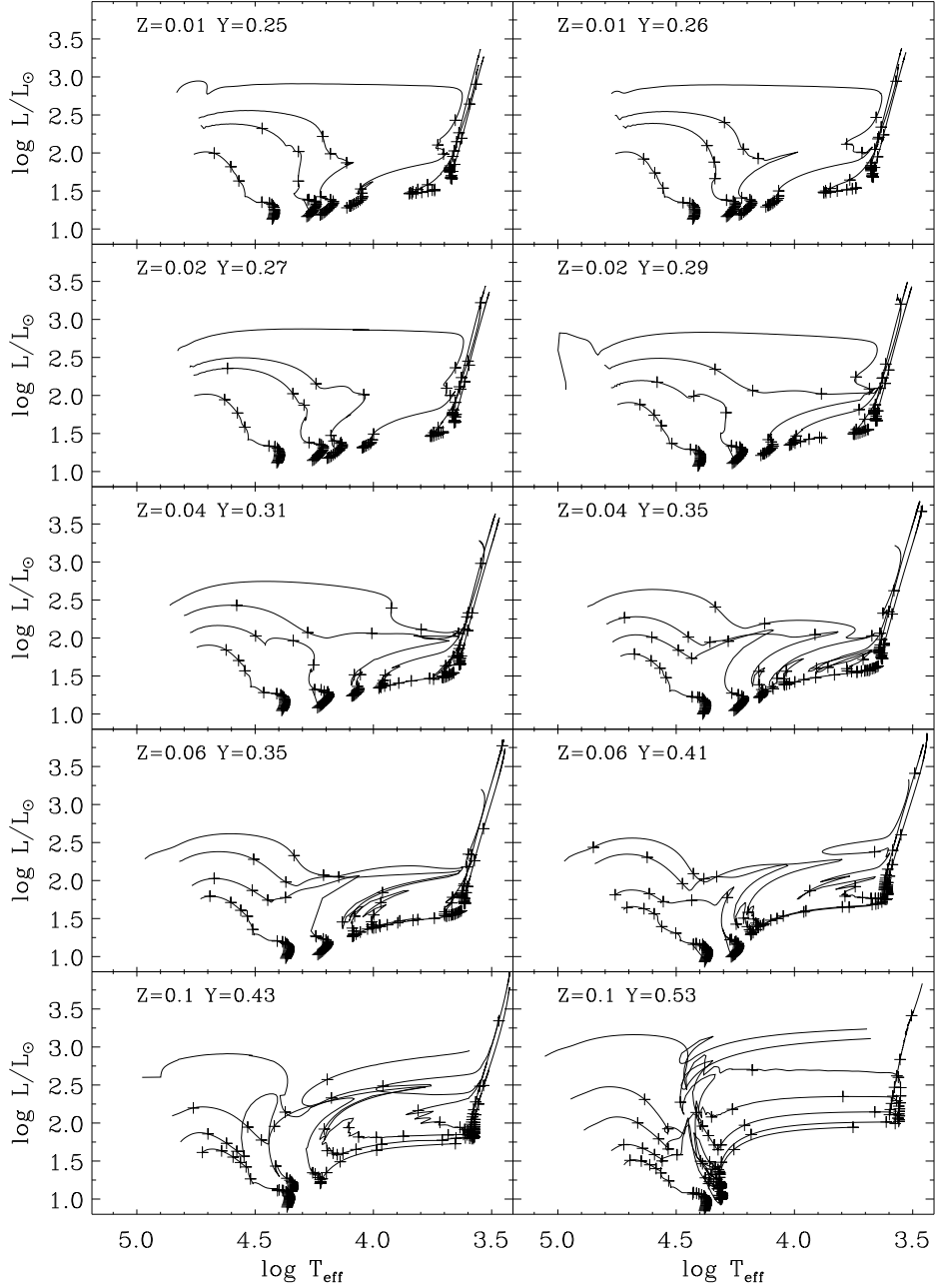}
\caption{Same as Figure 1, but for $Z \geq 0.01$. 
Left and right panels are for \DYDZ=2 \& 3, respectively. 
Note that more massive stars evolve into the UV bright phase rather than into
AGB as metallicity increases. This phenomenon is more
conspicuous for \DYDZ=3. \label{fig2}}
\end{figure}

\clearpage
\begin{figure}
\plotone{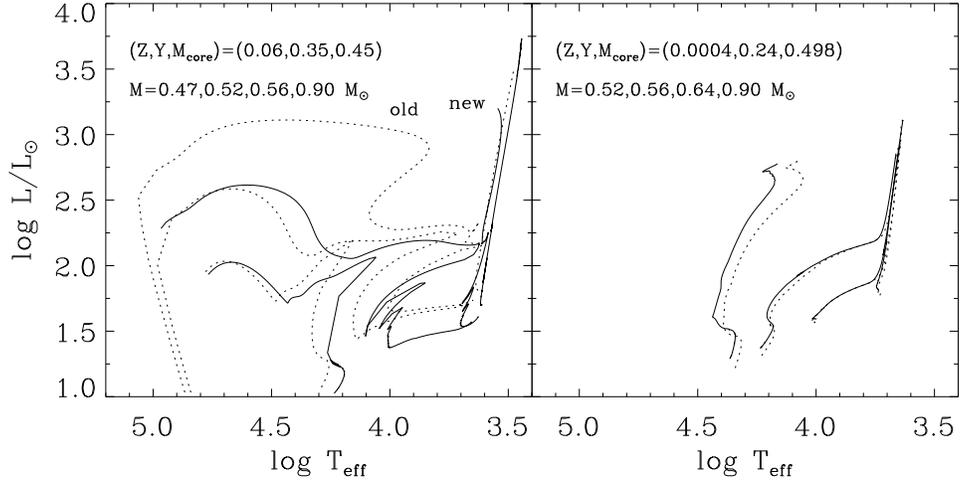}
\caption{Effects of improved physics on the evolutionary 
tracks. The solid and dotted lines are the models based on the improved and 
old physics, respectively. The effects are larger
on the metal-rich models that are sensitive to the change in opacities and
for the stars of $M \approx M_{tot}^{tr}$. For example, the new 
$M = 0.56 M_{\odot}$ model (a solid line designated by ``new'' in the left 
panel) does not become UV bright whereas the old model (dotted line with
``old'') does. \label{fig3}}
\end{figure}

\clearpage
\begin{figure}
\plotone{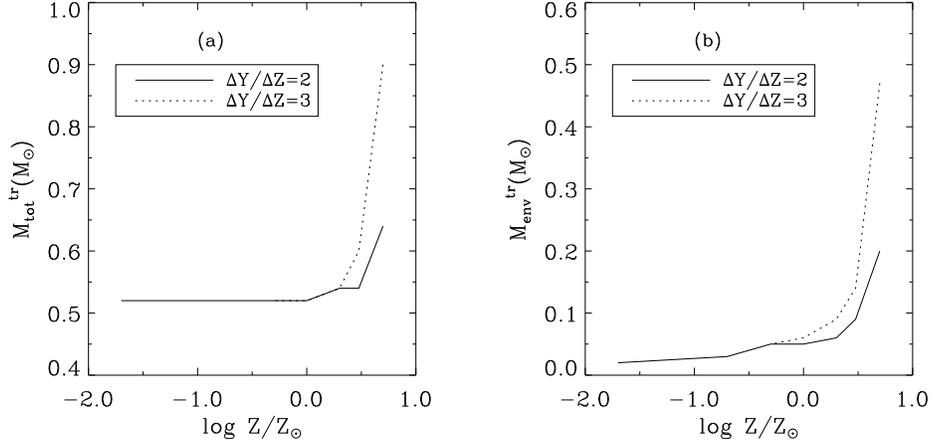}
\caption{ransition total mass $M_{tot}^{tr}$ (a) and 
transition envelope mass $M_{env}^{tr}$ (b) as a function of chemical 
composition. $M_{env}^{tr}$ is a monotonic function of chemical composition. 
\label{fig4}}
\end{figure}

\clearpage
\begin{figure}
\plotone{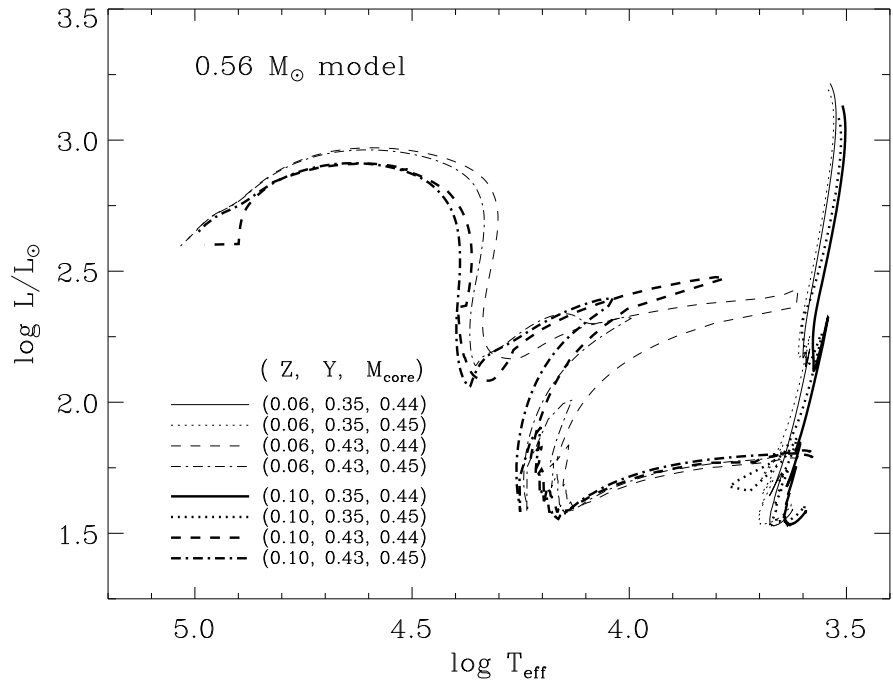}
\caption{Evolutionary tracks of the helium-burning stars of 
the same mass, but with different model parameters. Notice that the models 
with higher helium abundance become UV bright stars instead of asymptotic 
giant branch stars. \label{fig5}}
\end{figure}

\clearpage
\begin{figure}
\epsscale{0.8}
\plotone{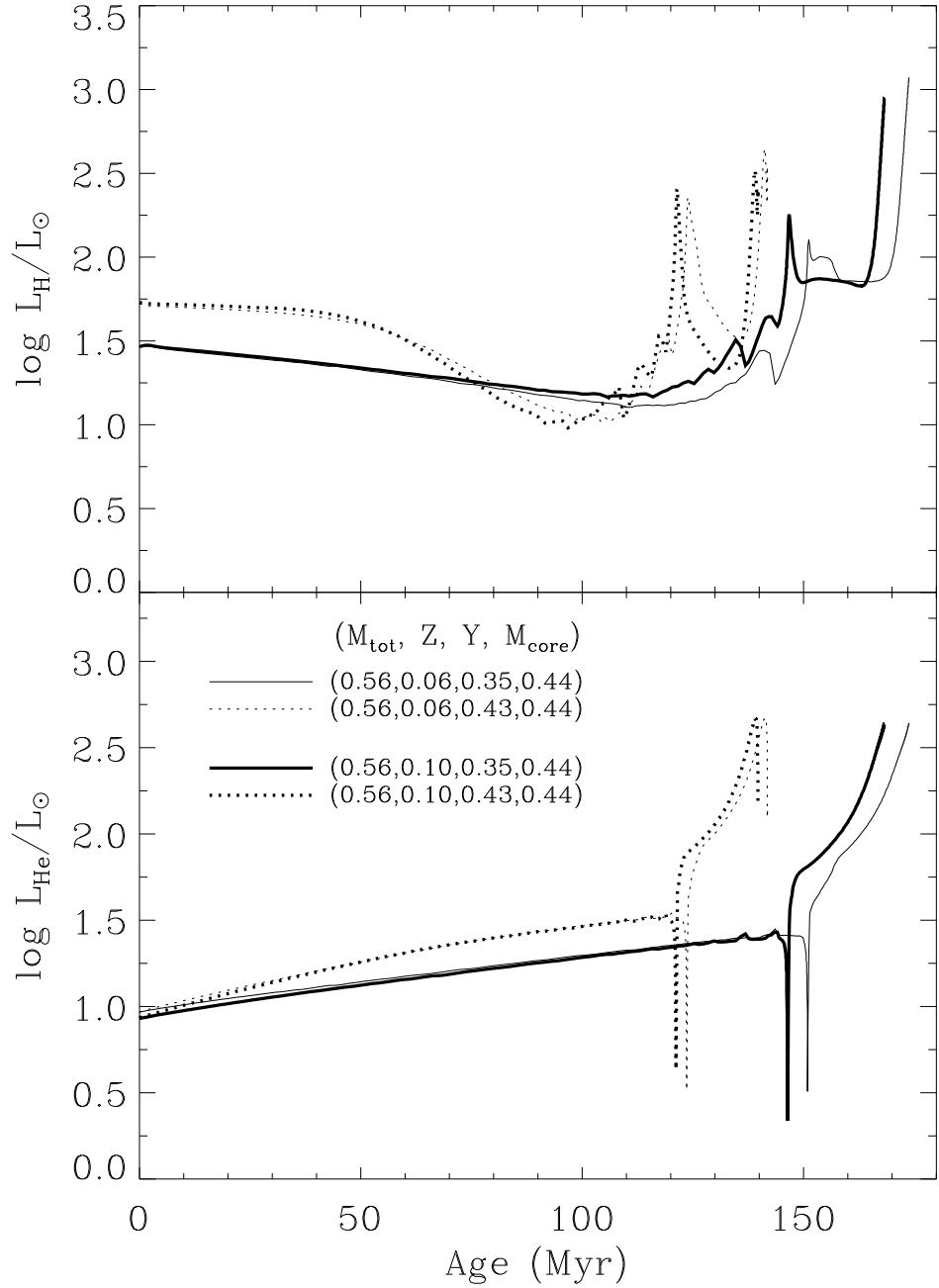}
\caption{Luminosity contribution from hydrogen and helium 
burning in the core 
helium-burning stars. Note that the hydrogen (CNO) luminosity in the stars 
with higher $Y$ is initially much higher than that in the stars with lower 
$Y$, while helium luminosity is less affected.\label{fig6}}
\end{figure}

\clearpage
\begin{figure}
\plotone{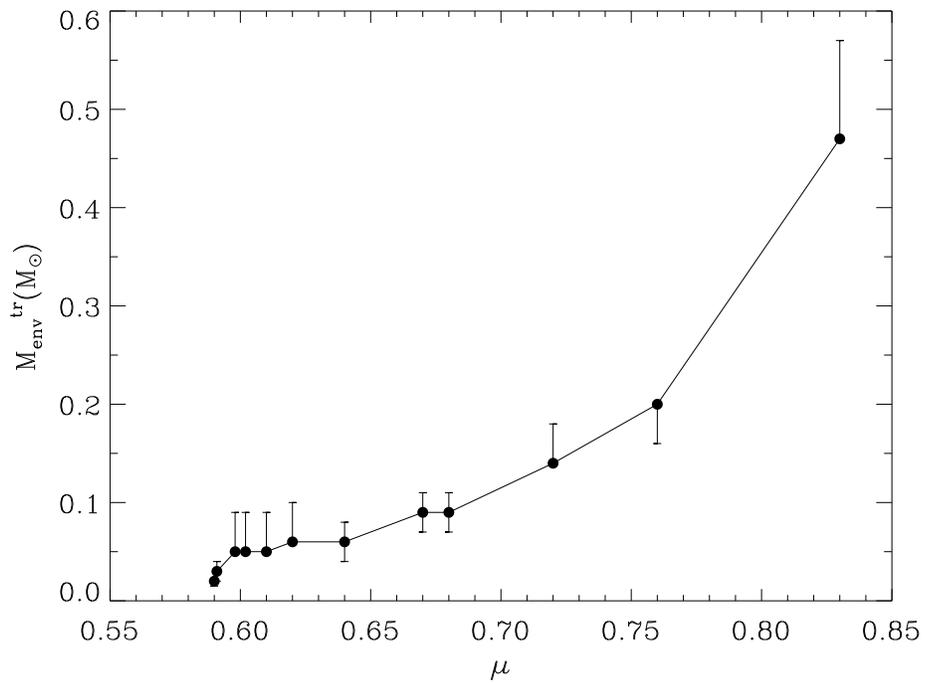}
\caption{Transition envelope mass $M_{env}^{tr}$ as a 
function of mean molecular 
weight $\mu$. A strong correlation exists. Error bars are from the uncertainty
in the $M_{env}^{tr}$ determination due to the lack of fine grid for mass in
the model construction. \label{fig7}}
\end{figure}
\end{document}